\documentclass{article}

\def\xxinput#1{\input#1}

\xxinput{vsolj02.sty}

\usepackage[dvipdfmx]{graphicx}
\usepackage[comma,colon]{natbib}

\usepackage[OT2,T1]{fontenc}

\def\cite{\citealt}
\setcitestyle{aysep={}}

\newcounter{author}
\def\altaffilmark#1{$^{#1}$}
\def\altaffiltext#1{$^{#1}$\,}

\def\authorcount#1#2{{\refstepcounter{author}\label{#1}
                     \altaffiltext{\ref{#1}}{#2}}}

\begin{document}

\begin{center}

\title{The 2022 active state of the AM CVn star NSV 1440}

\author{
        Taichi~Kato\altaffilmark{\ref{affil:Kyoto}},
        Rod~Stubbings\altaffilmark{\ref{affil:Stubbings}}
}

\authorcount{affil:Kyoto}{
     Department of Astronomy, Kyoto University, Sakyo-ku,
     Kyoto 606-8502, Japan \\
     \textit{tkato@kusastro.kyoto-u.ac.jp}
}

\authorcount{affil:Stubbings}{
     Tetoora Observatory, 2643 Warragul-Korumburra Road, Tetoora Road,
     Victoria 3821, Australia \\
     \textit{stubbo@dcsi.net.au}
}

\end{center}

\begin{abstract}
\xxinput{abst.inc}
\end{abstract}

   NSV 1440 was discovered as a variable star (BV 1025)
with a photographic range of 12.6 to fainter than 15.0
\citep{kni67nsv1440}.  The variability of this object was
not studied further and was listed as a suspected variable
in \citet{NSV}.  In 2015, the All-Sky Automated Survey for Supernovae
(ASAS-SN, \cite{ASASSN}) detected the transient ASASSN-15sz
(K. Stanek, vsnet-alert 19288)\footnote{
  $<$http://ooruri.kusastro.kyoto-u.ac.jp/mailarchive/vsnet-alert/19288$>$.
},
which was soon identified with NSV 1440
(D. Denisenko, vsnet-alert 19289)\footnote{
  $<$http://ooruri.kusastro.kyoto-u.ac.jp/mailarchive/vsnet-alert/19289$>$.
}.
Subsequent time-resolved photometry immediately detected
short-period variations (vsnet-alert 19291\footnote{
  $<$http://ooruri.kusastro.kyoto-u.ac.jp/mailarchive/vsnet-alert/19291$>$.
} and
vsnet-alert 19293\footnote{
  $<$http://ooruri.kusastro.kyoto-u.ac.jp/mailarchive/vsnet-alert/19293$>$.
}), followed by rapid fading
(vsnet-alert 19298)\footnote{
  $<$http://ooruri.kusastro.kyoto-u.ac.jp/mailarchive/vsnet-alert/19298$>$.
}.
These variations were soon identified as early superhumps and
ordinary superhumps [see e.g., \citet{kat15wzsge}
for the nomenclature of superhumps in WZ Sge stars],
and the object was suggested to be the first AM CVn star
showing both early and ordinary superhumps
(vsnet-alert 19304)\footnote{
  $<$http://ooruri.kusastro.kyoto-u.ac.jp/mailarchive/vsnet-alert/19304$>$.
}.  This object showed multiple rebrightenings
(starting with the one reported in vsnet-alert 19328)\footnote{
  $<$http://ooruri.kusastro.kyoto-u.ac.jp/mailarchive/vsnet-alert/19328$>$.
}, which has been now identified as one of
the characteristic features of AM~CVn-type superoutbursts
(see e.g., \cite{kat21newAMCVn}).
Another very similar superoutburst with multiple rebrightenings
occurred in 2017 and \citet{iso19nsv1440} reported on these
two superoutbursts.  \citet{iso19nsv1440} measured the periods
of early and ordinary superhumps to be 0.0252329(49)~d
and 0.025679(20)~d, respectively.  The former is expected to be
very close to the orbital period ($P_{\rm orb}$).
\citet{iso19nsv1440} determined a mass ratio of $q$=0.045(2).

   Not much work was done since \citet{iso19nsv1440} for
this object, one of the reasons probably being the object
near the south celestial pole, which is inaccessible from
northern observers.  Although \citet{pic21amcvns} referred to
the presence of Transiting Exoplanet Survey Satellite (TESS)
\citep{ric15TESS}\footnote{
  $<$https://tess.mit.edu/observations/$>$.
  The full light-curve
  is available at the Mikulski Archive for Space Telescope
  (MAST, $<$http://archive.stsci.edu/$>$).
} observations of this object, nothing particular was discussed.
This is understandable, since (currently) public TESS data
only recorded quiescent parts and we could not detect
a periodic signal corresponding to $P_{\rm orb}$.

   One of the authors (RS) noticed a number of short outbursts
in 2022 by visual observations.  Using ASAS-SN observations and
Asteroid Terrestrial-impact Last Alert System
(ATLAS: \cite{ATLAS}) forced photometry \citep{shi21ALTASforced},
we found that this object was indeed brighter its ordinary
quiescence [$BP$=18.44 and $RP$=18.42 according to \citet{GaiaDR3}]
and the even reached $g$=16.5 between these outbursts.

\begin{figure*}
\begin{center}
\includegraphics[width=16cm]{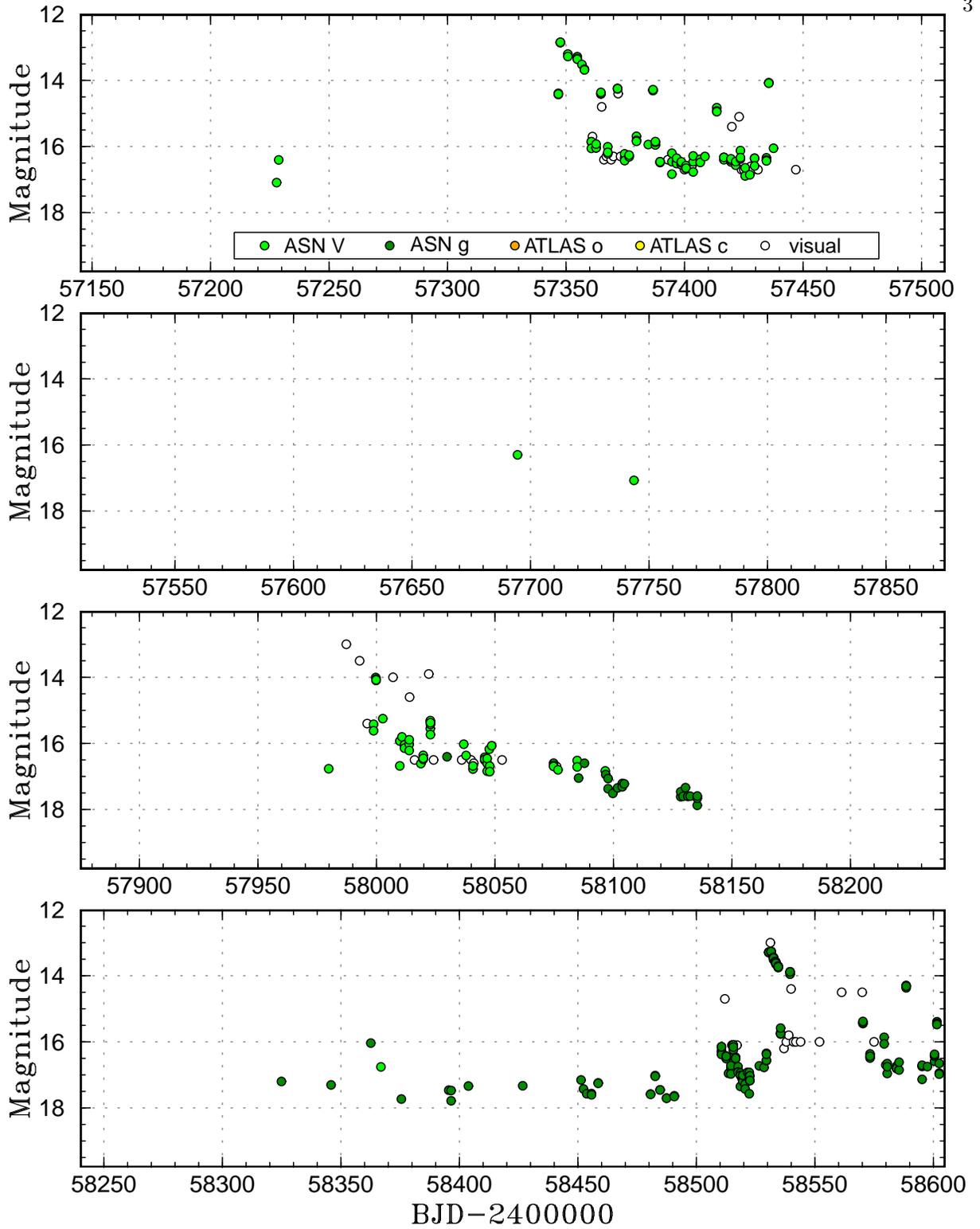}
\caption{
   Light curve of NSV 1440 in 2015--2019.
ASN and ``visual'' refer to ASAS-SN observations and
visual observations by RS, respectively.
Only positive detections are shown and all other ASAS-SN
observations were upper limits.  No other major outburst
was missed.
}
\label{fig:lc1}
\end{center}
\end{figure*}

\begin{figure*}
\begin{center}
\includegraphics[width=16cm]{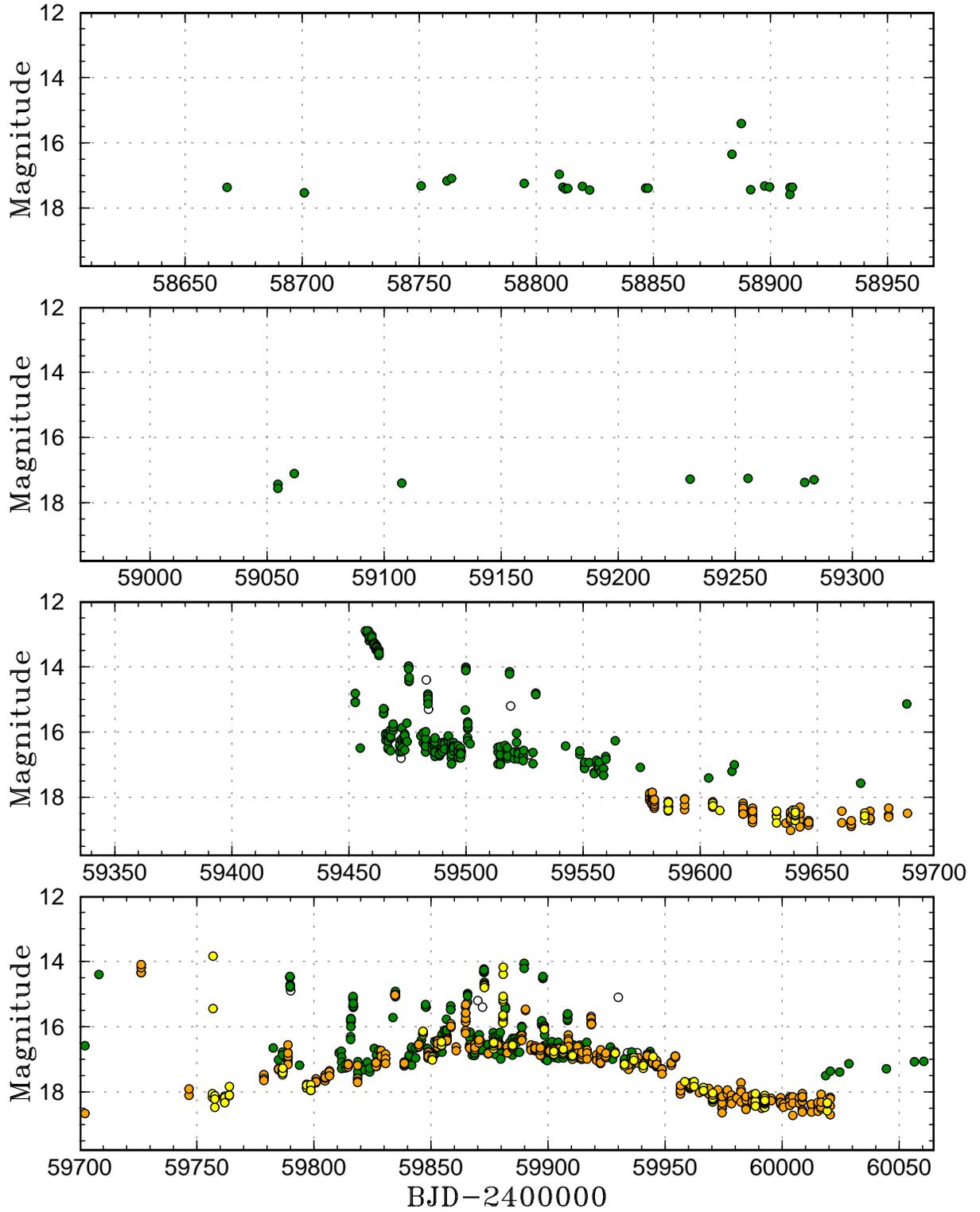}
\caption{
   Light curve of NSV 1440 in 2019--2023.
The symbols are the same as in figure \ref{fig:lc1}.
}
\label{fig:lc2}
\end{center}
\end{figure*}

\begin{figure*}
\begin{center}
\includegraphics[width=16cm]{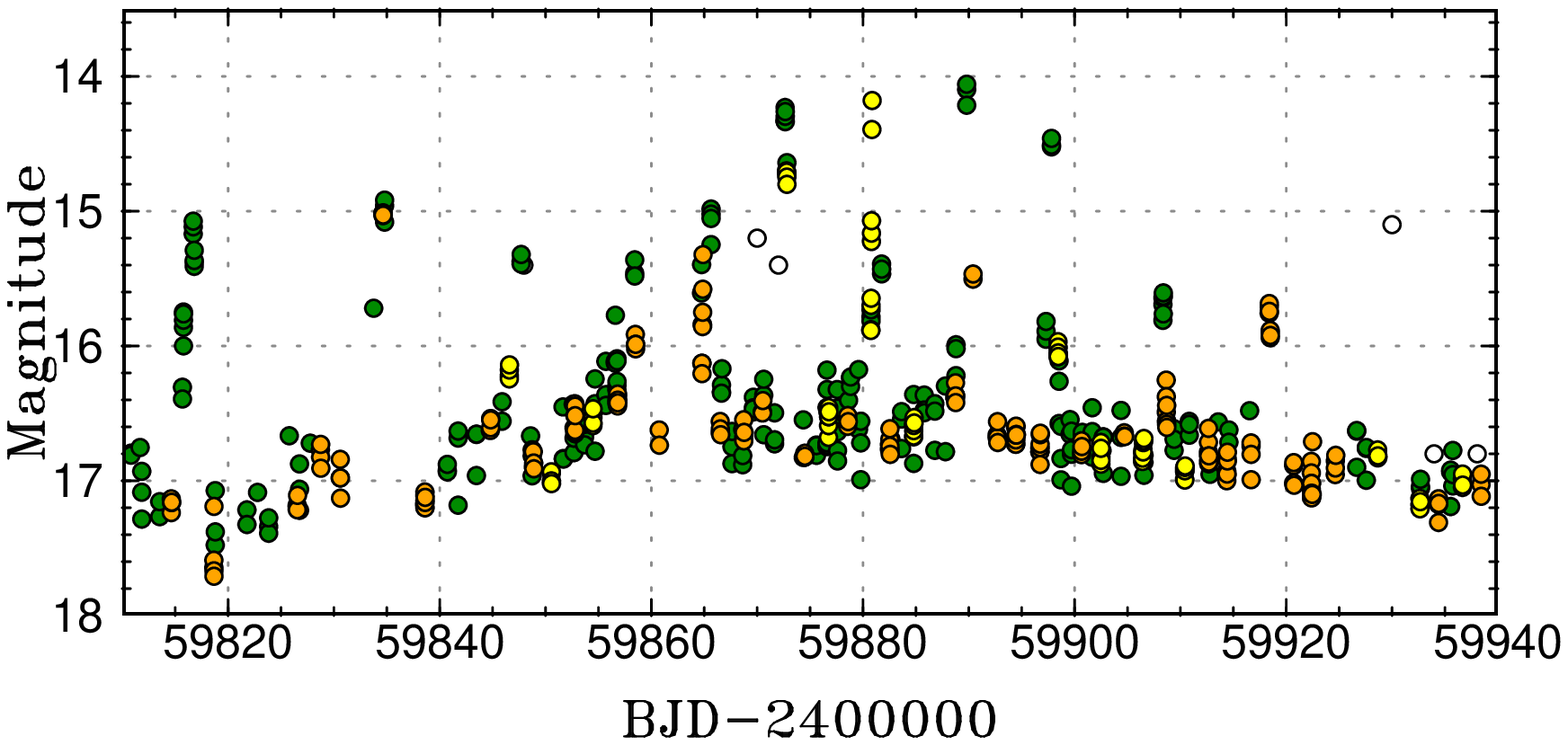}
\caption{
   Enlargement of the central part of the 2022 active state
in NSV 1440.
The symbols are the same as in figure \ref{fig:lc1}.
}
\label{fig:lc2022}
\end{center}
\end{figure*}

   The long-term light curves are shown in figures
\ref{fig:lc1} and \ref{fig:lc2}.  Only positive detections
are shown and all other ASAS-SN observations were
upper limits.  Although there are segments with only
a few points, a missed long outburst is excluded
from these ASAS-SN upper-limit observations.
Since the end of 2021, ATLAS observations became available
and the variation of the quiescent level and occurrence
of short outbursts are clearly shown in the third
and fourth panels of figure \ref{fig:lc2}.
An enlargement of the 2022 active state is shown in figure
\ref{fig:lc2022}.
We must note, however, isolated ASAS-SN points near
quiescence of these figures should be regarded as upper limits
rather than real values since they were incidentally positively
detected among other upper-limit observations due to
random errors.
A phenomenon similar to the 2022 active state was not
present at least between 2015 and 2021 based on observations
by the ASAS-SN and RS.

   Superoutbursts shown in these figures are
summarized in table \ref{tab:super}.  The dates are given
for optical peaks.  The 2019 superoutburst was preceded
by a precursor outburst.
All these outbursts had a short initial segment
(4--6~d) followed by complex rebrightenings
(both long and short ones depending on the superoutburst)
and a fading tail (see \cite{kat21newAMCVn}), which is
commonly seen after AM CVn-type superoutbursts.
Among these superoutbursts, the 2021 one appears to
have been the most powerful one in that it had the longest
duration (6~d) of the initial segment.  This is probably
a consequence of the long quiescent interval before
this superoutburst.  A larger mass should have been accumulated
in the disk before this superoutburst.

\begin{table*}
\caption{Superoutbursts in NSV 1440.}
\label{tab:super}
\begin{center}
\begin{tabular}{cccccc}
\hline
Year & Month & Day & BJD & Peak magnitude & Interval (d) \\
\hline
2015 & November & 21 & 2457350 & $V$=12.9 & -- \\
2017 & August   & 21 & 2457987 & $m_{\rm vis}$=13.0 & 637 \\
2019 & February & 16 & 2458531 & $m_{\rm vis}$=13.0 & 544 \\
2021 & August   & 30 & 2459457 & $m_{\rm vis}$=12.9 & 926 \\
\hline
\end{tabular}
\end{center}
\end{table*}

   After this 2021 superoutburst, the quiescent level
strongly varied, associated by frequent occurrence
of short outbursts as initially reported by RS.
There are two possibilities for this phenomenon
(and possibly a combination of them):
\begin{itemize}
\item The mass transfer-rate ($\dot{M}_{\rm tr}$) varied and
outbursts occurred frequently when $\dot{M}_{\rm tr}$ was high.
\item The quiescent viscosity varied and
outbursts occurred frequently when the viscosity was high,
if there was sufficient mass in the disk.
\end{itemize}
Both possibilities appear to naturally explain the observed
phenomenon.  AM CVn can have variable $\dot{M}$ as
in the recent example of a standstill in CR Boo
\citep{kat23crboo} and the variation in $\dot{M}_{\rm tr}$
could to be a viable explanation.
The variation in the quiescent viscosity also appears
to be promising, and has been proposed to explain
post-superoutburst rebrightenings in hydrogen-rich systems
(WZ Sge stars) \citep{osa01egcnc,mey15suumareb}.
The same explanation would apply to a helium disk,
and since the helium disk requires a higher temperature
to partially ionize \citep{tsu97amcvn,kot12amcvnoutburst},
a small change in the quiescent viscosity would
produce a more prominent effect than in hydrogen-rich
dwarf novae.
These possibilities may be distinguished by observing
the next superoutburst.  If $\dot{M}_{\rm tr}$ indeed
increased, the mass of the disk increased more rapidly
than in ordinary quiescence and the next superoutburst
is expected to occur earlier than expected or to be
a powerful one.
If the quiescent viscosity increased and $\dot{M}_{\rm tr}$
remained the same, the mass in the disk should have been
drained more quickly than in ordinary quiescence and
the next superoutburst is expected to occur later.
It is a pity that this object does not show
orbital modulations even in TESS data and it was
impossible to detect a possible enhancement of
the hot spot resulting from increased $\dot{M}_{\rm tr}$.

   It may be worth noting that this active state in 2022
was very similar in brightness and duration to
the post-superoutburst fading tail of the 2021 superoutburst.
The mechanism of fading tails of AM CVn stars is
still poorly understood.
In hydrogen-rich WZ Sge stars, very long-duration
(reaching a few years) fading tails have been considered
to arise from the cooling white dwarf
\citep{szk98alcom,god04wzsge,lon04wzsge,pir05wzsgeWD,god06wzsgeHST}.
This explanation, however, cannot be directly applicable
to fading tails of AM CVn-type superoutbursts at least
for two reasons:
(1) the mass accumulated or accreted in AM CVn-type dwarf novae
is much smaller and (2) the fading tail ends abruptly
(see the third panel of figure \ref{fig:lc2}), in contrast
to smooth fading expected for a cooling white dwarf.
The same feature of abrupt fading was also present in
the 2022 active phase: a sudden drop was present around
BJD 2459955.  These features suggest rapid switching between
high and low states, either by variable $\dot{M}_{\rm tr}$
or quiescent viscosity.
If the mechanism of the 2022 active phase is clarified,
it will also help understanding the yet unsolved mechanism
of the fading tail and associated rebrightenings in
AM CVn stars (see e.g., \cite{kat21newAMCVn,riv22asassn21au}).
This might also provide a clue in understanding the exceptional
hydrogen-rich dwarf nova V3101 Cyg with a long-lasting 
post-superoutburst phase with increased outburst activity
\citep{tam20v3101cyg,ham21DNrebv3101cyg}.

   We should make a comment on long-lasting faint ``superoutbursts''
claimed in the past.  \citet{riv20j0807} reported a faint
``superoutburst'' lasting more than a year in SDSS J080710.33$+$485259.6
with a photometric period of 53.3(3)~min \citep{kup19j0807atel12558}.
This ``superoutburst'' was very unusual in its morphology
and the peak brightness (18.2~mag) was only $\sim$2~mag
above the quiescence.  \citet{riv21j1137} reported a similar
0.5--1~mag (according to their paper; the amplitude was
1.5~mag according to the ATLAS data) long-lasting brightening
in SDSS J113732.32$+$405458.3 with $P_{\rm orb}$ of 59.6$\pm$2.7~min
\citep{car14amcvn}.  Based on these two ``outbursts'',
\citet{riv22asassn21au} suggested the presence of two types
of superoutbursts in long-$P_{\rm orb}$ AM CVn stars.
One type has a large amplitude and a short duration,
and it follows the tendency expected from
the disk-instability model.
This type is what we referred to superoutbursts here.
The other type with a small amplitude and a long duration
appears to be similar to the active state of NSV 1440 in 2022,
although short outburst on them were absent in these two
objects claimed in the past.
The peak $M_V$ of the long ``outburst'' in
SDSS J113732.32$+$405458.3 was $+$10.9 using the Gaia parallax
\citep{GaiaDR3} and the ATLAS data.  This value is close to
the peak $M_V$ = $+$8.9 in the active state of NSV 1440 in 2022.
The long ``outburst'' in SDSS J113732.32$+$405458.3 was probably
a phenomenon similar to the active state in NSV 1440 rather than
a true outburst considering the presence of genuine superoutbursts
in NSV 1440 with much larger amplitudes.  Although the parallax of
SDSS J080710$+$485259 was not determined well, the peak $M_V$
is expected to be similar to that of SDSS J113732.32$+$405458.3
considering the similar amplitude.
These active states should not be called outbursts or superoutbursts,
and they should not be discussed as outbursts in the context
of disk instability.

\section*{Acknowledgements}

This work was supported by JSPS KAKENHI Grant Number 21K03616.
The authors are grateful to the ATLAS and ASAS-SN teams
for making their data available to the public.
We are also grateful to Naoto Kojiguchi for helping downloading
TESS data.

This work has made use of data from the Asteroid Terrestrial-impact
Last Alert System (ATLAS) project.
The ATLAS project is primarily funded to search for
near earth asteroids through NASA grants NN12AR55G, 80NSSC18K0284,
and 80NSSC18K1575; byproducts of the NEO search include images and
catalogs from the survey area. This work was partially funded by
Kepler/K2 grant J1944/80NSSC19K0112 and HST GO-15889, and STFC
grants ST/T000198/1 and ST/S006109/1. The ATLAS science products
have been made possible through the contributions of the University
of Hawaii Institute for Astronomy, the Queen's University Belfast, 
the Space Telescope Science Institute, the South African Astronomical
Observatory, and The Millennium Institute of Astrophysics (MAS), Chile.

\section*{List of objects in this paper}
\xxinput{objlist.inc}

\section*{References}

We provide two forms of the references section (for ADS
and as published) so that the references can be easily
incorporated into ADS.

\newcommand{\noop}[1]{}\newcommand{\hyphalt}{-}

\renewcommand\refname{\textbf{References (for ADS)}}

\xxinput{nsv1440aph.bbl}

\renewcommand\refname{\textbf{References (as published)}}

\xxinput{nsv1440.bbl.vsolj}

\end{document}